\documentclass[aps,pra,twocolumn,floatfix]{revtex4}
\usepackage{epsfig}
\usepackage{graphicx}
\usepackage{dcolumn}
\usepackage{color}
\usepackage{longtable}
\usepackage{amsthm,amsmath}
\usepackage{color}
\usepackage{natbib}

\begin{document}

\title{Accurate determination of energy levels, hyperfine structure constants, lifetimes and dipole polarizabilities of triply ionized tin isotopes}

\author{Mandeep Kaur$^1$, Rishabh Nakra$^1$, Bindiya Arora$^1$\footnote{bindiya.phy@gndu.ac.in}, Cheng-Bin Li$^2$\footnote{cbli@wipm.ac.cn} and B. K. Sahoo$^{3}$\footnote{bijaya@prl.res.in}}
\affiliation{$^1$Department of Physics, Guru Nanak Dev University, Amritsar, Punjab-143005, India}
\affiliation{$^2$State Key Laboratory of Magnetic Resonance and Atomic and Molecular Physics, Wuhan Institute of Physics and Mathematics, Chinese Academy of Sciences, Wuhan 430071, China}
\affiliation{$^3$Atomic, Molecular and Optical Physics Division, Physical Research Laboratory, Navrangpura, Ahmedabad-380009, India}

\date{Received date; Accepted date}

\begin{abstract}
We have investigated energies, magnetic dipole hyperfine structure constants ($A_{hyf}$) and electric dipole (E1) matrix elements of a number of low-lying
states of the triply ionized tin (Sn$^{3+}$) by employing relativistic coupled-cluster theory. Contributions from the Breit interaction and 
lower-order quantum electrodynamics (QED) effects in determination of above quantities are also given explicitly. These higher-order relativistic 
effects are found to be important for accurate evaluation of energies, while QED contributions are seen to be contributing significantly to the 
determination of $A_{hyf}$ values. 
Our theoretical results for  energies are in agreement with one of the measurements but show significant differences 
for some states with another measurement. Reported $A_{hyf}$ will be useful in guiding measurements of hyperfine levels in the stable isotopes of 
Sn$^{3+}$. The calculated E1 matrix elements are further used to estimate oscillator strengths, transition probabilities and dipole polarizabilities 
($\alpha$) of many states. Large discrepancies between present results and previous calculations of oscillator strengths and transition 
probabilities are observed for a number of states. The estimated $\alpha$ values will be useful for carrying out high precision measurements using 
Sn$^{3+}$ ion in future experiments. 
\end{abstract}

\maketitle

\section{Introduction}
The spectra of medium to highly charged ions have aroused considerable interest in recent years for their applications in fundamental physics 
such as in search for variation of fundamental constants~\cite{berengut},  development of high precision optical frequency standards 
\cite{kozlov,yu2016,yu2018}, establishing very long-baseline interferometery for telescope array synchronization, development of extremely sensitive quantum 
tools for geodesy~\cite{Bloom, hinkley}, astronomy~\cite{sterling,11} and plasma physics \cite{su,22,2}. Despite the
well-understood nature of the force that binds these charged ions, highly accurate calculations of their properties are difficult and relatively 
sparse. In the present work,  we have considered medium charged Sn$^{3+}$  ion from Ag-like isoelectronic sequence for theoretical 
investigation of various spectroscopic properties. For this ion both Coulomb interactions and relativistic effects will be equally important to 
obtain its properties accurately. We have carried out calculations by including these interactions in the relativistic coupled-cluster 
(RCC) theory. 

Theoretical 
calculations and measurements of energies of ground and some of the low-lying states of Sn$^{3+}$ are available in literature \cite{scheers,gar,glow,ryab,moore,safr}.  Scheers et al.~\cite{scheers} obtained optical spectra of Sn$^{3+}$ from a laser produced plasma. In the same work they also made relativistic Fock-space coupled-cluster (FSCC) calculations of the measured energy levels. Safronova et. 
al.~\cite{safronova2014} presented Ag-like isoelectronic sequences and showed that they satisfy the criteria for experimental exploration in many 
fields of physics. Safronova et. al.~\cite{safr} used relativistic many-body perturbation theory (RMBPT method) to determine energies and lifetimes of the $4F_j$, $5P_j$ 
and $5D_j$ states in Sn$^{3+}$. 
Only a few spectroscopic studies of some of the low lying states of Sn$^{3+}$ have been carried out \cite{safr,biswas,cheng}.  Safronova et al.~\cite{safr} had determined oscillator strengths for the $5S_j-5P_{j'}$, $5P_j-5D_{j'}$, $4F_j-5D_{j'}$ and 
$4F_j-5G_{j'}$ transitions using RMBPT method. Biswas et al.~\cite{biswas} had estimated transition properties for 33 lines of Sn$^{3+}$ using the relativistic 
coupled-cluster (RCC) theory. In their analysis corrections due to Briet and quantum electrodynamics (QED) effects such as the electron 
self-energy and vacuum polarization interactions were omitted. The oscillator strengths of a few transitions are given by Cheng and Kim~\cite{cheng} using multi-configuration 
Dirac-Fock (MCDF) method.  There have been a few measurements of the lifetimes of some of the states of 
Sn$^{3+}$~\cite{andersen,kernahan,pin} using the beam foil technique.  Lifetime calculation of few excited states of Sn$^{3+}$ have been made by Cheng et al.~\cite{cheng} using relativistic Hartree-Fock method and Pinnington et al.~\cite{pin} using Coulomb approximation. 

In this work, we intend to investigate roles of electron correlation 
effects and higher-order relativistic corrections using RCC theory for accurate calculation of various properties in Sn$^{3+}$. For this purpose, we
present energies and magnetic dipole hyperfine structure constants ($A_{hyf}$) of the $nS_{1/2}$($n=5-9$), $nP_{1/2,3/2}$ ($n=5-9$), 
$nD_{3/2,5/2}$($n=5-9$) and $nF_{5/2,7/2}$($n=4,5$) states of Sn$^{3+}$. Accurate determination of $A_{hyf}$ values are very sensitive to the 
relativistic effects owing to its origin from atomic nucleus \cite{morrison}. We have also determined electric dipole (E1) matrix elements among 
the above states. Using these elements, we further determine transition probabilities and oscillator strengths for many transitions. Additionally, we estimate lifetimes and static dipole
polarizabilities of many states. A comparison between some of our theoretical calculations and other experimental and theoretical values available 
in the literature is also presented. Spectroscopic investigations of Sn ions carried out in this work have potential applications in laser produced plasmas (LPPs)~\cite{monchinsky, su, tomie, white},
future thermonuclear fusion reactors~\cite{3,4,5} and their discovery in various stellar and interstellar 
atmospheres~\cite{lunt, toole, vennes, 12, 7, proffitt, 9}. 
Our results on $A_{hyf}$ values and dipole polarizabilities will be useful for comprehensive understanding of 
roles of electron correlation effects and higher-order relativistic effects in their determination by comparing them with the experimental values 
when available in future. 
 In Secs. \ref{sec2}, \ref{sec3} and \ref{sec4}, we present theory and method of calculations, the results and discussion, and the conclusions subsequently. 

\begin{table*}
\caption{Energy levels of Sn$^{3+}$ (in cm$^{-1}$). The theoretical values obtained in this work using DHF and RCCSD methods are presented.  
The ``Total'' column lists the sum of the RCCSD calculations including Breit interaction ($\delta_{\rm{Breit}}$) and QED effects 
($\delta_{\rm{QED}}$). A comparison to theoretical values using relativistic many-body perturbation theory (RMBPT) calculations~\cite{safr} 
and Fock space coupled-cluster (FSCC) calculations~\cite{scheers} is listed. The experimental values obtained in Ref.~\cite{scheers},
~\cite{ryab} and ~\cite{moore} are compared. Uncertainties are presented in the parentheses whereas only  statistical uncertainty is presented in parentheses for the experimental results from  Ref.~\cite{ryab}\label{tab1}. }
\begin{ruledtabular}
\begin{tabular}{rrrrrrrrrr}

\multicolumn{1}{c}{$nl_j$} &
\multicolumn{1}{c}{ DHF} &
\multicolumn{1}{c}{ RCCSD} &
\multicolumn{1}{c}{$\delta_{\rm{Breit}}$} &
\multicolumn{1}{c}{$\delta_{\rm{QED}}$} &
\multicolumn{1}{c}{ Total} &
\multicolumn{1}{c}{RMBPT}~\cite{safr}&
\multicolumn{1}{c}{FSCC}~\cite{scheers}&
\multicolumn{1}{c}{Experiment}~\cite{scheers}&
\multicolumn{1}{c}{Experiment(others)}\\
\hline\\
$5S_{1/2}$ & -315810.897  & -328953.136 & 150.639 & 105.221& -328697(967) & -327453 &-328999 &-328908.4(3)& -328550(300)\cite{moore}\\
$5P_{1/2}$ & -249487.205  & -259271.768 & 200.228 & 14.946 & -259057(685) & -258188 &-259258 &            & -258986.1 \cite{moore}\\
$5P_{3/2}$ & -243519.185  & -252691.321 & 115.82  & -6.559 & -252582(653)  & -251717 &-252743 &           & -252477.7\cite{moore}\\
$5D_{3/2}$ & -159328.353  & -163254.896 & 17.706  & 0.349  & -163237(298) & -162915 & -163353 &-163604(1)& -163245.3\cite{moore}\\
$5D_{5/2}$ & -158629.079  & -162499.782 & -2.706  & -1.781 & -162504(276) & -162170 &-162617 &-163499(1)& -163139.2\cite{moore}\\
$6S_{1/2}$ & -151212.132  & -154796.302 & 45.723  & 30.269 & -154720(299)  &        &-154763 &-154769.6(5)& -154411.2\cite{moore}\\
$6P_{1/2}$ & -128067.360  & -131018.250 & 67.571  & 4.882  & -130946(225) &         &-130974 &-131057.8(6)& -130699.1\cite{moore}\\
$6P_{3/2}$ & -125965.710  & -128801.549 & 39.757  & -2.053 & -128764(216) &         &-128783 &-128878.3(5)& -128519.2\cite{moore}\\
$4F_{7/2}$ &  -115110.552 & -118084.917 & -13.779 & 1.113  & -118098(226) & -118035 &-118445 &-118650.2(6)& -118292.3\cite{moore}\\
$4F_{5/2}$ &  -115024.424 & -118014.829 & -12.51  & 1.056  & -118026(226) & -117959 &-118372 &-118590.5(7)& -118231.8\cite{moore}\\
$6D_{3/2}$ & -92462.213   & -92893.595  & 9.743   & 2.355  & -92881(988)  &         &-93828  &-94111(1)& -93754.3\cite{moore}\\
$6D_{5/2}$ & -92145.028   & -92585.920  & 0.817   & -1.149 & -92586(963)  &         &-93502  &-93779.7(3)& -93422.3\cite{moore}\\
$7S_{1/2}$ &  -89756.877  &  -91295.451 & 20.592  & 13.454 & -91261(132)  &         &-91079  &-91291(1)& -90934.3\cite{moore} \\
$7P_{1/2}$ &  -78777.021  &  -80132.272 & 32.201  & 2.362  & -80098(105)  &         &-79905  &-80173.0(3)&\\
$7P_{3/2}$ &  -77765.509  &  -79084.579 & 19.277  & -0.993 & -79066(102)  &         &-78766  &-79263.6(3)& \\
$5F_{7/2}$ & -74678.943   &  -76407.429 & -13.366 &-28.943 & -76450(182)  & -76472  &-76373  &-77055(2)&\\
$5F_{5/2}$ & -74622.998   &  -76425.892 & -10.095 & 18.336 & -76418(137) & -76428   &-76333  &-76745.8(3)&\\
$7D_{3/2}$ &  -60806.413  &  -61621.898 & 4.542   & 0.021  & -61617(62)  &         &-61524   &-61692.9(3)& -61653.4\cite{ryab}\\
$7D_{5/2}$ &  -60635.790  &  -61441.207 & -0.378  & -0.525 & -61442(61)   &         &-61352  &-61513.7(3)& -61514.3\cite{ryab}\\
$8S_{1/2}$ &  -59507.940  &  -60326.596 & 11.166  & 7.252  & -60308(72)  &         &-60104   &-60364.1(4)& -60366\cite{ryab}\\
$8P_{1/2}$ & -53344.583   &  -53532.865 & 17.488  & 1.006  & -53514(18)  &         &&&\\
$8P_{3/2}$ &  -52768.349  & -52944.746  & 10.353  & -0.718 & -52935(14)  &         &&&  \\
$8D_{3/2}$ &  -43085.314  &  -43561.413 & 2.697   & 0.164  & -43558(36)  &         &-43502   &-43643(1)&\\
$8D_{5/2}$ &  -42984.845  &  -43455.457 & -0.176  & -0.308 & -43456(36)  &         &-43402   &-43538(1)&\\
$9S_{1/2}$ &  -42125.075  &  -42649.840 & 7.229   & 4.679  & -42638(46)  &         &&& -42897\cite{ryab}\\
$9P_{1/2}$ &  -38088.328  & -38189.801  & 12.33   & 0.755  & -38177(10)  &         &&&\\
$9P_{3/2}$ &  -37677.634  & -37768.907  & 7.579   & -0.478 & -37762(8)  &         &&&\\
$9D_{3/2}$ &  -31940.746  &  -32190.898 & 1.442   & 0.006  & -32189(19)  &         &&&\\
$9D_{5/2}$ &  -31887.893  &  -32135.193 & -0.071  & -0.162 & -32135(19)  &         &&&\\

\end{tabular}
\end{ruledtabular}
\end{table*}

\begin{table*}[t]
\caption{Calculated $A_{hyf}/g_I$ values (in MHz) from the DHF and RCCSD methods are presented using the DC Hamiltonian. Corrections from the 
Breit interaction ($\delta_{\rm{Breit}}$) and QED effects ($\delta_{\rm{QED}}$) from the RCCSD method are also quoted. Rough estimations of uncertainties
from partial excitations to some of the final results are given in the parentheses. Combining these calculations with the $g_I$ values as $-1.83766$, $-2.00214$ and $-2.09456$ 
of $^{115}$Sn, $^{117}$Sn and $^{119}$Sn isotopes, respectively, $A_{hyf}$ values of many low-lying states of $^{115,117,119}$Sn$^{3+}$ ions 
are listed.}
\begin{ruledtabular}
\begin{tabular}{rrrrrrrrr}
 & \multicolumn{5}{c}{$A_{hyf}/g_I$ values} & \multicolumn{3}{c}{$A_{hyf}$ constants} \\
 \cline{2-6} \cline{7-9} \\
\multicolumn{1}{c}{$nl_j$} &
\multicolumn{1}{c}{ DHF} &
\multicolumn{1}{c}{ RCCSD} &
\multicolumn{1}{c}{$\delta_{\rm{Breit}}$} &
\multicolumn{1}{c}{$\delta_{\rm{QED}}$} &
\multicolumn{1}{c}{ Total} &
\multicolumn{1}{c}{$^{115}$Sn$^{3+}$} &
\multicolumn{1}{c}{$^{117}$Sn$^{3+}$}&
\multicolumn{1}{c}{$^{119}$Sn$^{3+}$} \\
\hline\\
$5S_{1/2}$ & 19513.74  & 23081.31 & 10.18 & $-194.48$ & 22897(90) & $-42077(165)$ & $-45843(180)$ & $-47959(190)$ \\ 
$5P_{1/2}$ & 4121.52  & 4968.22  & $-6.31$ & $-7.22$ & 4955(15) & $-9106(28)$ & $-9921(30)$ & $-10379(31)$ \\
$5P_{3/2}$ &  636.21 & 824.43 & 1.33 & $-0.44$ & 825(4) & $-1516(7)$ & $-1652(8)$ & $-1728(8)$ \\
$5D_{3/2}$ & 149.73  & 215.04 & 0.83 & $-0.06$ & 216.0 & $-397.0$ & $-433.0$ & $-453.0$ \\
$5D_{5/2}$ &  62.62 &  89.70 & 0.42 & 0.01 & 90.0 & $-165.0$ & $-180.0$ & $-189.0$ \\
$6S_{1/2}$ & 5900.49  &  6706.81 & 5.06 &  $-54.50$ & 6657(20) & $-12233(37)$ & $-13328(75)$ & $-13944(78)$ \\
$6P_{1/2}$ &  1429.30 &  1646.58 & $-1.03$ & $-2.25$ & 1643(5) & $-3019(9)$ & $-3290(10)$ & $-3441(11)$  \\
$6P_{3/2}$ &  225.35 &  284.06 & 0.54 & $-0.18$ & 284(3) & $-522(6)$ & $-569(6)$ & $-595(6)$ \\
$4F_{7/2}$ & 5.32  & $-8.56$ & $-0.04$ & 0.08 & $-9.0$ & 17.0 & 18.0 & 19.0 \\
$4F_{5/2}$ & 9.53 & 11.03  &  0.02 & $-0.09$ & 11.0 & $-20.0$ & $-22.0$ & $-23.0$ \\
$6D_{3/2}$ &  68.38  & 76.17 & 0.35 & $-0.10$ & 76.0 & $-140.0$ & $-152.0$ & $-159.0$ \\
$6D_{5/2}$ &  28.61  & 32.22 & 0.08 &  $-0.02$ & 32.0 & $-59.0$ & $-64.0$ & $-67.0$ \\
$7S_{1/2}$ & 2664.14  & 2987.10 & 2.57 & $-24.03$ & 2966(8) & $-5451(15)$ & $-5938(16)$ & $-6213(17)$ \\
$7P_{1/2}$ &  685.19  & 773.54 & $-0.25$ & $-0.98$ & 772(3) & $-1419(6)$ & $-1546(6)$ & $-1617(6)$ \\
$7P_{3/2}$ &  108.82  &  136.70 & 0.29 & $-0.58$ & 136(2) & $-250(4)$ & $-272(4)$ & $-285(4)$ \\
$5F_{7/2}$ &  4.19  &  $-5.92$ & $\sim 0$ & $\sim 0$ & $-6.0$ & 11.0 & 12.0 & 13.0 \\
$5F_{5/2}$ & 7.54   & 14.76 & 0.06 & $\sim 0$ & 15.0 & $-28.0$ & $-30.0$ & $-31.0$ \\ 
$7D_{3/2}$ & 36.90  & 48.36 & 0.19 & 0.01 & 49.0 & $-90.0$ & $-98.0$ & $-103.0$ \\  
$7D_{5/2}$ & 15.45 & 22.12 & 0.09 & $-0.01$ & 22.0 & $-40.0$ & $-44.0$ & $-46.0$ \\
$8S_{1/2}$ & 1445.69 & 1610.77 & 1.49 & $-12.88$ & 1599(4) & $-2938(8)$ & $-3201(9)$ & $-3345(9)$ \\
$8P_{1/2}$ & 390.04 & 415.16 & $-0.30$ & $-0.59$ & 414.0 & $-761.0$ & $-829.0$ & $-867.0$ \\
$8P_{3/2}$ & 62.02 & 70.94 & 0.12 & $-0.03$ & 71.0 & $-131.0$ & $-142.0$ & $-149.0$ \\
$8D_{3/2}$ & 21.76 & 28.35 & 0.11 & $\sim 0$ & 29.0 & $-53.0$ & $-58.0$ & $-61.0$ \\
$8D_{5/2}$ & 9.13 & 12.83 & 0.05 & $\sim 0$ & 13.0 & $-24.0$ & $-26.0$ & $-27.0$ \\
$9S_{1/2}$ & 934.36 & 1039.00 & 0.99 & $-8.29$ & 1032(2) & $-1897(4)$ & $-2066(4)$ & $-2162(4)$ \\
$9P_{1/2}$ & 279.19 & 296.38 & $-2.26$ & $-0.42$ & 294.0 & $-540.0$ & $-589.0$ & $-616.0$ \\
$9P_{3/2}$ &  43.97 & 49.33 & 0.34 & $-0.02$ & 50.0 & $-92.0$ & $-100.0$ & $-105.0$ \\
$9D_{3/2}$ & 11.46 & 14.97 & 0.06 & $\sim 0$ & 15.0 & $-28.0$ & $-30.0$ & $-31.0$ \\
$9D_{5/2}$ & 4.81 & 6.73 & 0.03 &  $\sim 0$ & 7.0 & $-13.0$ & $-14.0$ & $-15.0$ \\
\end{tabular}
\end{ruledtabular}
\label{tabAhyf}
\end{table*}

\section{Theory and Method of Calculations} \label{sec2}

We use the RCC theory with the singles and doubles excitations approximation (RCCSD method) 
(e.g. see Refs. ~\cite{PhysRevA.90.050503,PhysRevA.92.052511,PhysRevA.91.042507})  to determine the wave 
functions of the ground and excited states of the Sn$^{3+}$ ion. In this approach, we first obtain the Dirac-Hartree-Fock (DHF) wave function 
($|\Phi_0\rangle$) for the closed core of Sn$^{4+}$ then append the respective valence orbital ($v$) of the ground or excited state as 
$\vert \Phi_v \rangle= a_v^{\dagger}|\Phi_0\rangle$ to define the DHF wave function of Sn$^{3+}$ (also known as V$^{N-1}$ potential). Considering 
this DHF wave function as starting point, the exact atomic wave function ($|\Psi_v\rangle$) is determined by expressing it in the RCC theory as
\begin{equation}
|\Psi_v \rangle = e^T\{1+S_v\}|\Phi_v\rangle,
\end{equation}
where the RCC operators $T$ and $S_v$ are responsible for carrying out excitations of the core, and core and valence electrons respectively from the DHF 
wave functions due to the correlation effects. In the RCCSD method, these RCC operators can be given in the second quantization notations as
\begin{eqnarray}
 T= \sum_{a,p} \eta_a^p a_p^{\dagger} a_a + \frac{1}{4} \sum_{a,b,p,q}\eta_{ab}^{pq} a_p^{\dagger} a_q^{\dagger} a_b a_a  
\end{eqnarray}
and 
\begin{eqnarray}
 S_v= \sum_{p\ne v} \eta_v^p a_p^{\dagger} a_v + \frac{1}{2} \sum_{p,q,a} \eta_{va}^{pq} a_p^{\dagger} a_q^{\dagger} a_a a_v  ,
\end{eqnarray}
where $a,b$ and $p,q$ indices represent the occupied and unoccupied orbitals, respectively, and $\eta$ are the corresponding excitation 
amplitudes. These amplitudes are solved by using the following equations
\begin{eqnarray}
  \langle \Phi_0^K | \bar{H}_N | \Phi_0 \rangle = \delta_{K,0} (E_0-E_{DHF}) , 
  \label{eqt0} 
\end{eqnarray} 
and
\begin{eqnarray}
 \langle \Phi_v^L | \bar{H}_N \{ 1 + S_v \} | \Phi_v \rangle = (E_v-E_0) \langle \Phi_v^L | \{ \delta_{L,0} + S_v \} | \Phi_v \rangle  , \ \ \ \
 \label{eqamp}
\end{eqnarray}
where $\bar{H}_N \equiv e^{-T}He^T$ with the atomic Hamiltonian $H$ and subscript $N$ means normal order form with respect to the reference 
state $|\Phi_0 \rangle$. The superscripts $K$ and $L$ indicate $K^{th}$ and $L^{th}$ excited determinants with respect to $|\Phi_0 \rangle$ 
and $|\Phi_v \rangle$, respectively. Here, $E_{DHF}$ and $E_0$ are the DHF and total energies of the closed core, and $E_v$ is the total 
energy of the considered state of Sn$^{3+}$ containing the valence orbital $v$. Therefore, evaluation of $E_v-E_0$ will give the electron affinity of 
the corresponding valence orbital $v$. 

For the calculations, we consider first the Dirac-Coulomb (DC) Hamiltonian given by
\begin{eqnarray}\label{eq:DHB}
H^{DC} &=& \sum_i \left [c\mbox{\boldmath$\alpha$}_i\cdot \textbf{p}_i+(\beta_i-1)c^2+V_n(r_i)\right] +\sum_{i,j>i}\frac{1}{r_{ij}}, \ \ \ \
\end{eqnarray}
where $c$ is the speed of light, $\mbox{\boldmath$\alpha$}$ and $\beta$ are the usual Dirac matrices, $\textbf{p}_i$ is the single particle momentum operator, $V_n(r_i)$ denotes the nuclear potential,
and $\frac{1}{r_{ij}}$ represents the Coulomb potential between the electrons located at the $i^{th}$ and $j^{th}$ positions. 

We investigate the
Breit interaction contribution by including the following potential in the atomic Hamiltonian
\begin{eqnarray}\label{eq:DHB}
V^B &=& - \sum_{j>i}\frac{[\mbox{\boldmath$\alpha$}_i\cdot\mbox{\boldmath$\alpha$}_j+
(\mbox{\boldmath$\alpha$}_i\cdot\mathbf{\hat{r}_{ij}})(\mbox{\boldmath$\alpha$}_j\cdot\mathbf{\hat{r}_{ij}})]}{2r_{ij}} ,
\end{eqnarray}
where $\mathbf{\hat{r}_{ij}}$ is the unit vector along $\mathbf{r_{ij}}$.

\begin{table*}[t]
\caption{Transition rates ($A_{ik}$) with the power of 10 in brackets (in s$^{-1}$), absorption oscillator strengths $f_{ki}$ (in a.u.) and  
transition wavelengths $\lambda$ (in \AA) for transition from upper level $i$ to lower level $k$ are presented. The values of the oscillator 
strengths reported by Biswas et. al. ~\cite{biswas} using the RCCSD method, and Safronova et. al. ~\cite{safr} using the RMBPT method are also
given along with other literature values.}
\begin{ruledtabular}
\begin{tabular}{l r r r r r r c}
\multicolumn{2}{c}{Transition}& &&\multicolumn{4}{c}{$f_{ki}$}\\
\cline {1-2}  \cline{5-8}
\multicolumn{1}{c}{$i$ level} & \multicolumn{1}{c}{$k$ level} & \multicolumn{1}{c}{$\lambda$} &\multicolumn{1}{c}{$A_{ik}$} &
Present & Ref.~\cite{biswas}  & Ref.~\cite{safr} &Others \\
\hline 
  5$P_{1/2}$&  5$S_{1/2}$&    1437.527&  8.2355(8) &   0.25514  & 0.259   & 0.2489& 0.258~\cite{glow}, 0.243~\cite{gar}\\
  6$P_{1/2}$&  5$S_{1/2}$&     505.431&  2.8324(8) &   0.01085  & 0.01   & \\
  5$P_{3/2}$&  5$S_{1/2}$&    1314.539 & 1.0871(9) &   0.56327 &  0.572  & 0.5508& 0.567~\cite{glow} , 0.538~\cite{gar}\\
  6$P_{3/2}$&  5$S_{1/2}$&     499.923  & 1.4790(8) &   0.01108 &  0.01  & \\
  6$S_{1/2}$&  5$P_{1/2}$&     956.252 & 1.1655(9) &   0.15978 & 0.163  &  &0.165~\cite{gar} \\
  6$P_{1/2}$&  6$S_{1/2}$&    4217.256 & 1.6517(8) &   0.44041 & 0.445 &\\
  6$S_{1/2}$&  5$P_{3/2}$&    1019.716&  2.3132(9) &   0.18030 & 0.182 & &0.185~\cite{gar},0.180~\cite{pin}\\
  6$P_{3/2}$&  6$S_{1/2}$&    3862.197&  2.1257(8) &   0.95074  & 0.96\\
  7$S_{1/2}$&  5$P_{1/2}$&     595.055&  4.8927(8) &   0.02597 & 0.025 \\
  7$S_{1/2}$&  6$P_{1/2}$&    2514.787&  2.7268(8) &   0.25853 & 0.257  \\
  7$S_{1/2}$&  5$P_{3/2}$&     619.029&  9.4341(8) &   0.02710 & 1.354  \\
  7$S_{1/2}$&  6$P_{3/2}$&    2660.643& 5.4323(8) &   0.28826 & 0.285 \\
  5$D_{3/2}$&  5$P_{1/2}$&    1044.487&  2.9842(9) &   0.97617 & 0.986  & 0.9577 & 0.972 ~\cite{gar} \\
  6$D_{3/2}$&  5$P_{1/2}$&     605.210&   6.0840(8) &   0.06682 & 0.036  \\
  5$D_{3/2}$&  5$P_{3/2}$&    1120.669&  5.2344(8) &   0.09856 & 0.111 &  0.0968&  0.095~\cite{mig}, 0.088~\cite{gar}\\
  6$D_{3/2}$&  5$P_{3/2}$&     630.027&   9.0177(7) &   0.00537 & 0.661 \\
  5$D_{5/2}$&  5$P_{3/2}$&    1119.338&  3.1362(9) &   0.88364 & 1.005  & 0.8736 &0.885~\cite{gar}\\
  6$D_{5/2}$&  5$P_{3/2}$&     628.712&   5.7941(8) &   0.05150 & 5.945 \\
  6$P_{1/2}$&  5$D_{3/2}$&    3072.555&  3.1727(8) &   0.22452 & 0.231  & \\
  6$D_{3/2}$&  6$P_{1/2}$&    2706.741&   5.8312(8) &   1.28097 & 1.360 \\
  6$P_{3/2}$&  5$D_{3/2}$&    2879.678&  3.5903(7) &   0.04463 & 0.046 \\
  6$D_{3/2}$&  6$P_{3/2}$&    2876.464&   1.0631(8) &   0.13187 & 0.139 \\
  6$P_{3/2}$&  5$D_{5/2}$&    2888.504& 3.2976(8) &   0.27499 & 0.278 \\
  6$D_{5/2}$&  6$P_{3/2}$&    2849.254&   6.4735(8) &   1.18182 & 1.256 \\
  4$F_{5/2}$&  5$D_{3/2}$&    2221.556&  8.0330(8) &   0.89154 &0.914 &  0.8751 & 1.036~\cite{cheng}  \\
  4$F_{5/2}$&  5$D_{5/2}$&    2226.804&  5.7484(7) &   0.04273 & 0.044  & 0.0413 \\
  4$F_{7/2}$&  5$D_{5/2}$&    2229.809&  8.5841(8) &   0.85315 & 0.861  & 0.8251 & 0.977~\cite{cheng}  \\
  6$D_{3/2}$&  4$F_{5/2}$&    4085.385&   7.0379(7) &   0.11740 & 0.138 \\
  6$D_{5/2}$&  4$F_{5/2}$&    4030.714&  3.4000(6) &   0.00828 & 0.010  \\
  6$D_{5/2}$&  4$F_{7/2}$&    4020.909&   6.7996(7) &   0.12361 & 0.141 \\ 
\end{tabular}
\end{ruledtabular}
\label{tab2}
\end{table*}

Contributions from the QED effects are estimated by considering the lower-order vacuum polarization (VP) interaction ($V_{VP}$) 
and the self-energy (SE) interactions ($V_{SE}$). We account for $V_{VP}$ through the Uehling \cite{PhysRev.48.55} 
and Wichmann-Kroll potentials ($V_{VP}=V^{Uehl} + V^{WK}$) given by
\begin{eqnarray}
 \label{eq:uehl}
V^{Uehl}&=&- \frac{2}{3} \sum_i \frac{\alpha_e^2 }{r_i} \int_0^{\infty} dx \ x \ \rho_n(x)\int_1^{\infty}dt \sqrt{t^2-1} \nonumber \\
&& \times\left(\frac{1}{t^3}+\frac{1}{2t^5}\right)  \left [ e^{-2ct|r_i-x|} - e^{-2ct(r_i+x)} \right ]\ \ 
\end{eqnarray}
and
\begin{eqnarray}
 V^{WK} = \sum_i \frac{0.368 Z^2}{9 \pi c^3 (1+(1.62 c r_i )^4) } \rho_n(r_i),
\end{eqnarray}
respectively, with the electron density over the nucleus as $\rho_n(r)$ and the atomic number of the system as $Z$.

The SE contribution $V_{SE}$ is estimated by including two parts as
\begin{eqnarray}
V_{SE}^{ef}&=& - A_l \sum_i \frac{2 \pi Z \alpha_e^3 }{r_i} I_1^{ef}(r_i) + B_l \sum_i \frac{\alpha_e }{ r_i} I_2^{ef}(r_i) \ \ \
\end{eqnarray}
known as effective electric form factor part and
\begin{eqnarray}
V_{SE}^{mg}&=& \sum_k \frac{i\alpha_e^3}{4} \mbox{\boldmath$\gamma$} \cdot \mbox{\boldmath$\nabla$}_k \frac{1}{r_k} \int_0^{\infty} dx \ x \ \rho_n(x)
\int_1^{\infty} dt \frac{1}{t^3 \sqrt{t^2-1}} \nonumber \\
\times && \left [ e^{-2ct|r_k-x|} - e^{-2ct(r_k+x)} - 2ct \left (r_k+x-|r_k-x| \right ) \right ], \nonumber \\
\end{eqnarray}
known as effective magnetic form factor part. In the above expressions, we use \cite{Ginges-PRA-2016} 
\begin{eqnarray}
A_l= \begin{cases} 0.074+0.35Z \alpha_e \ \text{for} \ l=0,1 \\  0.056+0.05 Z \alpha_e + 0.195 Z^2 \alpha_e^2 \ \text{for} \ l=2  , \end{cases}
\end{eqnarray}
and
\begin{eqnarray}
B_l = \begin{cases} 1.071-1.97x^2 -2.128 x^3+0.169 x^4  \ \text{for} \ l=0,1 \\
     0 \ \text{for} \ l \ge 2 .  \end{cases}   
\end{eqnarray}
The integrals are given by
\begin{eqnarray}
I_1^{ef}(r) =  \int_0^{\infty} dx \ x \ \rho_n(x) [ (Z |r-x+1)| e^{-Z|r-x|} \nonumber \\  - (Z(r+x)+1) e^{-2ct(r+x)}  ] \ \ \ \ \ \ 
\end{eqnarray}
and
\begin{eqnarray}
 I_2^{ef}(r) &=& \int_0^{\infty} dx \ x \ \rho_n(x)  \int^{\infty}_1 dt \frac{1}{\sqrt{t^2-1}} \bigg \{ \left( 1-\frac{1}{2t^2} \right ) \nonumber \\
&\times& \left [ \ln(t^2-1)+4 \ln \left ( \frac{1}{Z \alpha_e} +\frac{1}{2} \right ) \right ]-\frac{3}{2}+\frac{1}{t^2} \big \}\nonumber \\
&\times& \{ \frac{\alpha_e}{t} \left [ e^{-2ct|r-x|} - e^{-2ct(r+x)} \right ] +2 r_A e^{2 r_A ct } \nonumber \\
&\times& \left [ E_1 (2ct (|r-x|+r_A)) - E_1 (2ct (r+x+r_A)) \right ] \bigg \} \nonumber \\
\end{eqnarray}
with the orbital quantum number $l$ of the system, $x=(Z-80)\alpha_e$, $r_A= 0.07 Z^2 \alpha_e^3$, and the exponential integral $E_1(r) = 
\int_r^{\infty} ds e^{-s}/s$. We have used the Fermi nuclear charge distribution in our calculations by defining
\begin{equation}
\rho_n(r)=\frac{\rho_0}{1+e^{\frac{r-b}{a}}},
\end{equation}
for the normalization factor $\rho_0$, the half-charge radius $b$, and $a=2.3/4(\ln3)$ is related to the skin thickness. We have determined $b$ using the relation
\begin{equation}
b=\sqrt{\frac{5}{3}r^2_{rms}-\frac{7}{3}a^2\pi^2},
\end{equation}
with the root mean square (rms) charge radius of the nucleus evaluated by using the formula
\begin{equation}
r_{rms}=0.836A^{1/3}+0.570,
\end{equation}
in fm for the atomic mass $A$.

\begin{table*}[t]
\caption{Lifetimes for few excited states (in ns) calculated in the present work and from other available literature data. The numbers in
parentheses represent uncertainties. Theoretical calculations from  Ref.~\cite{safr} uses RMBPT, Ref.~\cite{biswas} uses the RCCSD
method, Ref.~\cite{cheng} uses relativistic Hartree-Fock method and Ref.~\cite{pin} uses Coulomb approximation  whereas experimental results are obtained using beam foil technique.}
\begin{ruledtabular}
\begin{tabular}{lccccccc}
\multicolumn{1}{c}{State} &
\multicolumn{1}{c}{Present} &
\multicolumn{1}{c}{Theory(others)} &
\multicolumn{1}{c}{Theory~\cite{pin}} &
\multicolumn{1}{c}{Experiment~\cite{kernahan}} & 
\multicolumn{1}{c}{Experiment~\cite{pin}}\\
\hline\\
$6S_{1/2}$ & 0.29  &   &0.4 &&  0.29(4)\\
$7S_{1/2}$ & 0.44  &  \\
$8S_{1/2}$ & 0.71  &   \\
$9S_{1/2}$ & 1.01   & \\
$5P_{1/2}$ & 1.21   & 1.2~\cite{biswas}, 0.95~\cite{cheng}, 1.26~\cite{safr}& 1.03&0.73(40)& 1.29(20)  \\
$5P_{3/2}$ & 0.92   & 0.9~\cite{biswas}, 0.74~\cite{cheng}, 0.95~\cite{safr}&  &0.93(23)& 0.81(15)\\
$6P_{1/2}$ & 1.31  &  &2.2 &&1.41(15) \\
$6P_{3/2}$ & 1.38   & &1.9  &&1.40(15)  \\
$7P_{1/2}$ & 1.90  & &  \\
$7P_{3/2}$ & 2.13  & &  \\
$8P_{1/2}$ & 4.89   & &   \\
$8P_{3/2}$ & 4.89   &  & \\
$5D_{3/2}$ & 0.29  & 0.28~\cite{biswas}, 0.26~\cite{cheng} ,0.34~\cite{safr}&0.3 &0.35(3)& 0.34(4)\\
$5D_{5/2}$ & 0.32   & 0.28~\cite{biswas}, 0.29~\cite{cheng}, 0.32~\cite{safr}&0.3 & 0.41(3)&0.45(5)\\
$6D_{3/2}$ & 0.69   && 0.7  &&1.20(25)& \\
$6D_{5/2}$ & 0.77   & &0.8  &&1.26(20) &\\
$4F_{5/2}$ &1.12&  1.0~\cite{cheng}, 1.13~\cite{safr}&1.1& 1.05(9)&1.25(20)\\
$4F_{7/2}$ & 1.17 & 1.1~\cite{cheng}, 1.38~\cite{safr}&1.1& 1.06(9)&1.30(20) 
\end{tabular}
\end{ruledtabular}
\label{tab3}
\end{table*}

\textbf{some more lines are required giving details how from the Hamiltonian wavefunctions are obtained.}
After obtaining the atomic wave functions, we evaluate reduced matrix elements of an operator between states $|\Psi_k\rangle$ and $|\Psi_i\rangle$ 
from the following RCC expression
\begin{equation}
\langle\Psi_k||{\bf O}||\Psi_i\rangle = \frac{\langle\Phi_k||\{1+S_k^{\dagger}\}\overline{\bf O}\{1+S_i\}||\Phi_i\rangle}{\sqrt{N_kN_i}},
\end{equation}
where $\overline{O}=e^{T^{\dagger}}Oe^T$ and $N_v = \langle \Phi_v|e^{T^{\dagger}}e^T+S_v^{\dagger}e^{T^{\dagger}}e^TS_v|\Phi_v\rangle$. For the 
evaluation of expectation value, both the states are taken to be same. The calculation procedures of these expressions are discussed in detail
elsewhere~\cite{PhysRevA.91.042507,PhysRevA.92.052511}.
Next we discuss in brief the method used for calculation of hyperfine structure constant, transition probability, lifetime and dipole polarizability.

\textit{Hyperfine structure constant ($A_{hyf}$):} For isotopes with nuclear spin $I=1/2$, the hyperfine levels of an atomic state can be expressed as 
\begin{eqnarray}
W_{F,J} &=& \frac{1}{2} A_{hf} [F(F+1)-I(I+1)-J(J+1)].
\end{eqnarray}
Here $F=I \oplus J$ with the total angular momentum  $J$ and magnetic dipole hyperfine structure constant  
\begin{eqnarray}
A_{hyf}&=& \mu_N g_I \frac {\langle |Psi||{\bf O}_{hyf}^{(1)}|||Psi\rangle}{\sqrt{J(J+1)(2J+1)}}, \label{eqna} 
\end{eqnarray}
where $\mu_N$ is the nuclear Bohr magneton, $g_I= \frac {\mu_I}{I}$ with the nuclear magnetic moment $\mu_I$ and  ${\bf O}_{hyf}^{(1)}$ is 
the electronic component of the spherical tensor describing hyperfine interaction in an atomic system.

\textit{Transition probability (A$_{ik}$) and lifetime ($\tau_i$):} The transition probabilities from upper level $i$ to lower level $k$ are obtained from the reduced matrix elements of electric dipole (E1)
operator ($D$) by using the following standard expression ~\cite{sobelman1979atomic} 
\begin{equation}
A_{ik}=\frac{2.02613 \times 10^{18}}{\lambda^3}\frac{|\langle\Psi_i||{\bf D}||\Psi_k\rangle|^2}{g_i}
\end{equation}
and the emission oscillator strengths  are given by~\cite{sobelman1979atomic} 
\begin{equation}
f_{ik}=-\frac{303.756}{g_i\lambda}|\langle\Psi_i||{\bf D}||\Psi_k\rangle|^2,\label{eq-os}
\end{equation}
where $\lambda$ is the transition wavelength expressed in \AA, $g_i$ is the degeneracy factor for the $i^{th}$ state, and 
$\langle\Psi_i||{\bf D}||\Psi_k\rangle$ are used in atomic units (a.u.) to obtain $A_{ik}$ in $s^{-1}$.  From Eq. (\ref{eq-os}), the absorption oscillator 
strengths can be deduced using the relation
\begin{equation}
f_{ki}=-\frac{g_i}{g_k}f_{ik}.
\end{equation}
The lifetime of the $i^{th}$ level is the inverse of the sum of the transition probabilities arising from all the low-lying levels and
is given as
\begin{equation}
\tau_i=\frac{1}{\sum_k A_{ik}}.
\end{equation}
It is to be noted here that we have neglected contributions from the forbidden channels to determine the lifetimes of the investigated atomic 
states as they are found to be extremely small.

\textit{Dipole polarizability ($\alpha$):} The static dipole polarizability ($\alpha_v^{(q)}$) of an atomic state with valence orbital $v$, which depends upon the dipole matrix 
elements and energies $E$, can be expressed as 
\begin{equation}
 \alpha_v^{(q)} =-2 \sum_{v\neq{k}} \frac{|\langle \Psi_v | D | \Psi_{k}\rangle |^2}{(E_v-E_k)}.
\end{equation}
Carrying out tensor decomposition, it can be divided into three parts as~\cite{PhysRevA.91.012705, PhysRevA.85.012506}
\begin{equation}
\alpha_v^{(q)} = \alpha_{v,c}^{(q)}+\alpha_{v,cv}^{(q)}+\alpha_{v,v}^{(q)},
\end{equation}
where $q=0$ and 2 represent scalar and tensor polarizabilities, respectively, and the notations $c, cv, $ and $v$ in the subscript 
correspond to core, core-valence, and valence correlations, respectively. The core contributions to the tensor component of polarizability  
is zero. The scalar component contributes to all the atomic states whereas the tensor component contributes to the states with total angular 
momentum $j>1/2$. It should be noted that $\alpha_{v,v}^{(q)}$ contributes the most in the evaluation of $\alpha_v^{(q)}$ in the considered 
states of Sn$^{3+}$. This contribution can be estimated to very high accuracy in the sum-over-states approach 
using the formula 
\begin{eqnarray}
\alpha_{v,v}^{(0)}  &=& 2 \sum_{k>N_c, k \ne v } W^{(0)}_{v}\frac{ |\langle \Psi_v|| D  || \Psi_k\rangle |^2} 
{(E_v - E_k)}, \nonumber \\
\end{eqnarray}
and 
\begin{eqnarray}
\alpha_{v,v}^{(2)}  &=& 2 \sum_{k>N_c, k \ne v } W^{(2)}_{v,k}\frac{  |\langle \Psi_v||  D  || \Psi_k\rangle |^2} 
{(E_v - E_k)}. \nonumber \\
\end{eqnarray}
with $N_c$ as the number of occupied orbitals and the coefficients as
\begin{eqnarray}
W_{v}^{(0)} &=&-\frac{1}{3(2J_v+1)}, \label{eqp0} 
\end{eqnarray}
and
\begin{eqnarray}
W_{v,k}^{(2)} &=&2\sqrt{\frac{5J_v(2J_v-1)}{6(J_v+1)(2J_v+3)(2J_v+1)}} \nonumber \\ 
& & \times (-1)^{J_v+J_k+1}
                                  \left\{ \begin{array}{ccc}
                                            J_v & 2 & J_v \\
                                            1 & J_k &1 
                                           \end{array}\right\}.         \label{eqp2}
\end{eqnarray} 

\begin{table*}[t]
\caption{The static scalar and tensor polarizabilities (in a.u.) for the ground state and few excited states. The RPA value
for the core contribution to the scalar polarizability $\alpha_{v,c}^{(0)}$ is estimated to be 2.264 a.u.. The core valence 
contributions are estimated to be approximately zero. $\alpha_{v}^{(q)}$ values include contributions from first few dominant 
transitions labeled as  ``Main($\alpha_{v,v}^{(q)}$)'',  higher excited states denoted as ``Tail($\alpha_{v,v}^{(i)}$)'' and 
core correlations  $\alpha_{v,c}^{(0)}$.}
\begin{ruledtabular}
\begin{tabular}{ccccccc}
\multicolumn{1}{c}{State} &
\multicolumn{1}{c}{Main($\alpha_{v,v}^{(0)}$)} &
\multicolumn{1}{c}{Tail($\alpha_{v,v}^{(0)}$)} &
\multicolumn{1}{c}{$\alpha_{v}^{(0)}$} &
\multicolumn{1}{c}{Main($\alpha_{v,v}^{(2)}$)} &
\multicolumn{1}{c}{Tail($\alpha_{v,v}^{(2)}$)} &
\multicolumn{1}{c}{$\alpha_{v}^{(2)}$}\\
\hline\\
$5S_{1/2}$ & 7.27      & 0.02   &   9.53  & - & - & -\\
$6S_{1/2}$ & 103.57  & 0.03   & 105.85 & - & - & -\\
$7S_{1/2}$ & 620.25  & 0.02  & 622.58 & - & - & -\\
$8S_{1/2}$ & 2078.72  & 0.23  & 2081.20 & - & - & -\\
$9S_{1/2}$ & 5866.38  & 0.96  & 5869.59 & - & - & -\\
$5P_{1/2}$ & 3.48      & 0.04   &   5.78 & - & - & -\\ 
$6P_{1/2}$ &  -3.56      & 0.254    &    -1.04  & - & - & -\\
$7P_{1/2}$ &   161.53     &  1.60    &    -157.66 & - & - & -\\
$5P_{3/2}$ & 4.670     & 0.03  & 6.97 & 0.77 & -0.026    & 0.74 \\
$6P_{3/2}$ & 10.25   & 0.23     &  12.74   & 20.58  & -0.15     & 20.43 \\
$7P_{3/2}$ &  -78.75  &   1.51   &  -74.97  & 145.39 &  0.96    & 144.44 \\
$5D_{3/2}$ & 30.32 & 0.52     &  33.11    & -11.05  & -0.11     &  -11.16  \\
$6D_{3/2}$ & 286.75 & 1.36    &  290.38   & -110.02 & -0.31     &  -110.33 \\
$5D_{5/2}$ & 29.30  & 0.56     & 32.12  & -13.82  & -0.18     &  -13.99 \\
$6D_{5/2}$ & 283.18 & 1.54    & 286.98  & -141.44 &-0.52  & -141.96 \\  
\end{tabular}
\end{ruledtabular}
\label{tab4}
\end{table*}

 In the above approach, we break the valence contribution into two parts: contributions from low-lying  $k$ states up to which we can 
determine $\langle\Psi_n||D||\Psi_k\rangle$ matrix elements using the RCCSD method and  experimental energies $E_i$s from Moore energy 
table~\cite{moore}, which are labeled as ``Main($\alpha_{n,v}^{(q)}$)'', and contributions from higher excited states,  
denoted as ``Tail($\alpha_{n,v}^{(i)}$)'', are estimated using the DHF method. Similarly, the core-valence contributions $\alpha_{n,cv}^{(0)}$ 
is also obtained using the DHF method using the expression
\begin{eqnarray}
\alpha_{v,cv}^{(0)}  &=& 2 \sum_{k}^{N_c} W_{v}^{(0)} \frac{  |\langle 
\Psi_v || {D} || \Psi_k \rangle_{DF} |^2} {(E^{DHF}_v - E^{DHF}_k)}, \nonumber \\
\end{eqnarray}
and
\begin{eqnarray}
\alpha_{v,cv}^{(2)}  &=& 2 \sum_{k}^{N_c} W_{v,k}^{(2)} \frac{ |\langle 
\Psi_v|| {D} ||\Psi_k \rangle_{DF} |^2} {(E^{DHF}_v - E^{DHF}_k)}. \nonumber \\
\end{eqnarray}
However, we have adopted relativistic random phase approximation (RPA method), as discussed in Ref.~\cite{PhysRevA.90.022511}, to evaluate  
$\alpha_{v,c}^{(0)}$ from the closed core. 

\section{Results and Discussion}\label{sec3}

In Table \ref{tab1}, we have provided the calculated energy values (in cm$^{-1}$) of many states of Sn$^{3+}$ from the DHF and RCCSD methods. 
The fourth and fifth columns, respectively, represent the corrections in energy values due to the Breit interaction and QED effects. The final 
results along with uncertainties are quoted as ``Total'' in the same table. The uncertainties are estimated by analyzing contributions from the
neglected triples excitations in the perturbative approach. From the present calculations, we see that the Briet interaction corrections to
energies in Sn$^{3+}$ are large as compared to QED cntributions; especially for the $S_{1/2}$, $P_{1/2}$ and $P_{3/2}$ states. 
The contributions from these two corrections, however, have comparable influence for the other states. 
This could be due to the fact that wave functions of these states 
penetrate less inside the nucleus.  A comparison of our theoretical energy values obtained using the RCCSD method is presented with other available theoretical 
calculations from the Fock-space coupled-cluster (FSCC) method \cite{scheers} and RMBPT~\cite{safr} in the same table. A reasonable agreement between our and other theoretical values is found.  We compare our results for energy levels of the considered states with the experimentally available energy 
data from Ref.\cite{scheers} in second last column of the same table. The values in the parantheses are the statistical uncertainties in their measurements 
so it will inappropriate to compare our results without actual experimental error bars.
On the other hand we notice that  the other experimental measurements by Ryabtsev et al.\cite{ryab} and  C. Moore~\cite{moore} endrose our theoretical calculations very well.

We present $A_{hyf}/g_I$ values in Table \ref{tabAhyf} for all the calculated states as mentioned above. We give these values from the DHF 
and RCCSD methods along with corrections from the Breit and QED interactions. We have also quoted rough estimation of uncertainties to 
some of these quantities from valence triple excitations adopting perturbative approach. These values may change significantly if full 
triples are taken into account. By combining the final results of $A_{hyf}/g_I$ with the $g_I$ values of three stable isotopes with $A=115$, 
117 and 119, of Sn we give $A_{hyf}$ values of the $^{115,117,119}$Sn$^{3+}$ ions in the same table. Here, we have neglected slight 
differences in the $A_{hyf/g_I}$ values due to different isotope masses. 
We propose experiments for measurement of hyperfine levels of $^{115,117,119}$Sn$^{3+}$ ions in future to validate our calculations.
These results can be further improved 
by considering  full triple excitations in the RCC theory.

Transition probabilities for as many as 155 transition lines are obtained for Sn$^{3+}$ in our present work. A few of these transitions for which comparison from previous literature is available are listed in column 4 of Table \ref{tab2}.
The transition probabilities for other transitions are tabulated in supplementary material. 
In the same table we also present the corresponding absorption oscillator strengths ($f_{ki}$) from the present work 
and previously reported theoretical values from Ref.~\cite{biswas}  in columns 5 and 6 respectively. Our calculated values of oscillator strengths are also compared with 
previously reported theoretical results by Safronova et. al~\cite{safr} and other literature in the last two columns of the same table. 
The oscillator strengths calculated by Biswas and co workers~\cite{biswas}, who also use RCCSD method, 
are generally in good agreement with our results but unusually large discrepancies were found  among few transitions. 
For instance, the oscillator strength values for $5D_{j}-5P_{3/2}$ transition by them differ from our calculations by approximately 12\% and 14\% respectively. A similar inconsistency is observed for  $6D_{j}-5P_{j'}$ oscillator strengths.
These discrepancies are attributed to the 
disagreement in  matrix elements values for these transitions (see table I in supplementary material for a comparison of matrix elements).  A close inspection
of the oscillator strengths of transitions $7S_{1/2}-5P_{1/2,3/2}$ and $7S_{1/2}-6P_{1/2,3/2}$ given by Biswas et al.~\cite{biswas} points out 
that results for the $7S_{1/2}-6P_{3/2}$ transition are not correct because of the following simple reason. Assuming that radial component of the 
wavefunctions between the $5P_{1/2}$ and $5P_{3/2}$ states, and also between the $6P_{1/2}$ and $6P_{3/2}$ states are almost similar, 
the E1 matrix elements between the $7S_{1/2}-5P_{1/2}$ and $7S_{1/2}-5P_{3/2}$ transitions and also between the $7S_{1/2}-6P_{1/2}$ and 
$7S_{1/2}-6P_{3/2}$ transitions should differ mainly because of the angular factors. As seen both Biswas et al.~\cite{biswas} and we have
obtained similar matrix elements between the $7S_{1/2}-6P_{1/2}$ and $7S_{1/2}-6P_{3/2}$ transitions; which differ by one and half times 
approximately. Therefore, a similar factor difference between the E1 matrix elements of the transitions $7S_{1/2}-5P_{1/2}$ and 
$7S_{1/2}-5P_{3/2}$ is expected.  We believe that our results are more reliable  and they match well with the other available
literature.
For the $5D_{j}-5P_{j'}$ transitions, the oscillator strengths are in agreement with the value given in Ref.~\cite{gar} which are 
evaluated using the core-potential in the Dirac-Fock (DF + CP) method. In Ref. \cite{glow} oscillator 
strengths for the $5P_j-5S_{1/2}$ transitions are calculated employing CI method and our values are in good agreement with their numbers.
We 
notice remarkable agreement of our results with the values calculated in Ref.~\cite{safr} using RMBPT.    Our oscillator strengths for the $4F_{5/2}-5D_{3/2}$ and 
$4F_{7/2}-5D_{5/2}$ transitions are not in agreement with the results from the calculations of Cheng and Kim \cite{cheng}. They used relativistic 
Hartree-Fock method whereas our calculations are based on the RCC method which includes correlation corrections to all orders. Our 
oscillator strength for the $6S_{1/2}-5P_{3/2}$ transition is very close to the experimental result of Pinnington et al~\cite{pin} which is 
measured using beam foil technique.

In Table \ref{tab3}, estimation of lifetimes for the ground and few excited states along with a comparison with other available literature is 
presented. Theoretical calculations for lifetimes of various states are available in Refs. \cite{biswas,cheng, safr, pin}. In Ref.~\cite{biswas} authors use the CCSD method, whereas RMBPT and DF + CP methods 
are employed for calculations in  Ref.~\cite{safr}  and Ref.~\cite{cheng} respectively.  Theoretical 
calculations from Ref.~\cite{pin} are obtained assuming LS coupling and Coulomb-approximation radial wave functions. In general Coulomb 
approximation is only strictly valid for highly excited states with non-penetrating wave functions. Hence, it is not a match for 
our sophisticated calculations using the CCSD method. Our calculations match very well with these 
theoretical results.  
Experimental lifetime measurements for some states of Sn$^{3+}$ using beam foil technique are available in literature \cite{kernahan,pin}. Our calculations mostly show agreement with the experimental  results within the experimental uncertainties except for discrepancies at few places.  We notice discrepancies between our 
lifetime calculations and measurements from Pennington~\cite{pin} for $5D$ and $6D$ states.
Similarly, our results 
for the $5P_{1/2}$ and $5D$ states are not consistent with the measurements in Ref.~\cite{kernahan}, but they match well with other 
theoretical and experimental investigations. Therefore, it calls for more 
theoretical and experimental investigations for the lifetimes of these  states in Sn$^{3+}$.  

In Table \ref{tab4}, calculated values of the static scalar dipole polarizabilities for the $nS_{1/2}$ (n=5-9), $nP_{1/2, 3/2}$ (n=5-7) and 
$nD_{3/2,5/2}$ (n = 5, 6) are listed. In the same table the values of static tensor dipole polarizabilities for the $nP_{3/2}$ (n=5-7) and 
$nD_{3/2,5/2}$ (n = 5, 6), are also tabulated. The dominant ``Main'' contributions to the valence correlation for the scalar and 
tensor dipole polarizabilities are presented along with the ``Tail'' parts. The core contribution has been calculated using RPA 
and is found to be 2.264 a.u.. The contributions of valence core correlations are realized to be very small and thus, they have been excluded
from the table. It is found in our calculations that Sn$^{3+}$ in its ground state will not response much to the electric field as shown by a 
small value of static scalar polarizability ($\alpha_v^{(0)}=9.53$ a.u.). This small $\alpha_v^{(0)}$ value is owing to very large energy 
differences between the ground and $5P$ states leading to very less contribution to the polarizability from the primary 
$5S_{1/2}-5P_{1/2,3/2}$ transitions.

\section{Conclusion} \label{sec4}

In summary, theoretical results of energies, magnetic dipole hyperfine structure constants ($A_{hyf}$) and electric dipole (E1) matrix
elements of many low-lying states of the Sn$^{3+}$ are presented. Transition probabilities and oscillator strengths of 155 spectral lines 
arising from the $nS_{1/2} (n=5-9), nP_{1/2,3/2} (n=5-9), nD_{3/2,5/2} (n=5-9)$ and $nF_{5/2,7/2}$ (n=4,5) states along with radiative 
lifetimes for the 18 levels and static dipole polarizabilities of the 15 states have been determined. These values were obtained by 
employing relativistic couple-cluster theory with singles and doubles approximation. The estimated transition probabilities, oscillator
strengths, and radiative lifetimes are generally found to be in good agreements with the available experimental data.   The reported 
polarizability results for Sn$^{3+}$ can be useful in estimating systematics for carrying out high precision spectroscopic measurements 
in this ion.

\section*{Acknowledgements}

The work of B.A. is supported by DST-SERB Grant No. EMR/2016/001228. B.K.S. would like to acknowledge the use of Vikram-100 HPC cluster 
facility of Physical Research Laboratory, Ahmedabad, India.


\end{document}